\documentstyle[12pt]{article}
\newcommand{\de}{{\bf d}}
\newcommand{\tr}[1]{\,{\rm tr}\,#1\,}

\begin{document}
\title{
Bicovariant differential calculus on quantum groups
from Poisson Lie structures
\thanks
{\it Invited talk given by G.E.Arutyunov at the XXX Winter School of
Theoretical Physics
in Karpacz}}
\author{I.Ya.Aref'eva
\thanks{Steklov Mathematical Institute,
Vavilov 42, GSP-1, 117966, Moscow, Russia; arefeva@qft.mian.su}\\
G.E.Arutyunov
\thanks{Steklov Mathematical Institute,
Vavilov 42, GSP-1, 117966, Moscow, Russia; arut@qft.mian.su}\\and \\
P.B.Medvedev \thanks
{Institute of Theoretical and Experimental Physics,117259 Moscow, Russia}
}
\date {}
\maketitle
\begin{abstract}
The aim of this lecture is to give a pedagogical explanation of
the notion of a Poisson Lie structure on the external algebra
of a Poisson Lie group which was introduced in our previous papers.
Using this notion as a guide we construct
quantum external algebras on $SL_q(N)$ with proper (classical) dimension.
\end{abstract}

\section{Introduction}
The prime objective of this lecture is a pedagogical explanation of
our recent results \cite{AM},\cite{AAM}
on the problem of constructing the external algebra
on the quantum special linear group $SL_q(N)$.

Roughly speaking one can say that there are two main approaches to
develop differential calculi on quantum groups.

The first one uses the Connes approach \cite{Conn} to the noncommutative
geometry and it has been initiated
by Woronowicz \cite{Wor1}. In this approach
a bicovariant complex of differential forms and the
nilpotent operator $\de$
of exterior derivative obeying the Leibniz rule
play a fundamental role.

Another approach utilizes the  Faddeev idea \cite{FF}
that all objects in the theory of quantum groups
should appear naturally as the result of quantization
of appropriate Poisson brackets \cite{Dr}-\cite{Ta}.

Aside from the obvious theoretical meaning this approach reveals also
the practical significance. The point is that the Woronowicz bicovariant
differential calculus (BDC) is formulated in a rather abstract
axiomatic way and to develop its concrete realizations one needs
to incorporate some additional information specifying a quantum group \cite{J}.
For instance, the explicit description of
defining relations of a quantum external
algebra appears to be a nontrivial task \cite{Car}-\cite{WZ}. Another
complicated problem  is to check for these algebras the
Diamond Condition \cite{M}-\cite{Sudb}. Concerning $SL_q(N)$ one finds that the
bicovariant differential complex involves an {\em extra element} which has
no natural classical counterpart in $SL(N)$ \cite{Zum} and this is the
main problem that we are going to address in this lecture.

Generally speaking, there are two possibilities to attack the problem. The
first one consists in relaxing the Woronowicz BDC axioms. As it was
pointed out a couple of years ago by Faddeev \cite{Fadd} one can relax the
Leibniz rule assuming something like
\begin{equation}\label{lei}
\de (fh)=(\de f)h + Lf(\de h).
\end{equation}
and in a recent paper \cite{FP} this idea was successfully applied to
$SL_q(N)$ thereby confirming the quantization procedure for $SL_q(2)$
\cite{AAM}.

Another possibility is just the direct use of the semiclassical strategy
and the first step here is to endow the algebra of differential forms on a
classical matrix Lie group with a natural Poisson Lie structure.
Following Woronowicz we employ the bicovariance condition to be the main
characteristic property of this structure.
Note that in a quantum case the requirement of bicovariance
aside from the obvious geometrical meaning has a physical interpretation
\cite{AV}-\cite{ArA}.
At the second step, these Poisson Lie
structures when being quantized should give as the result
algebras of quantum differential forms.
The operator $\de $ defined on these algebras
in pure algebraic terms may not satisfy in general the standard requirements,
{\em
i.e.} the non-zero "anomaly" $L-1$ appears.
In the semiclassical limit the modified Leibniz rule results in a
{\em nondifferential}  Poisson Lie structure on the external algebra:
$$
\de \{f,h\}\neq\{\de f,h\}+\{f,\de h\}.
$$

The lecture consists of two parts. The first part deals with $GL(N)$ and
the second one with $SL(N)$. The first part explains the very notion of
a Poisson Lie structure on the external algebra of a Lie group and
following \cite{AM} we give the complete classification of these
structures for $GL(N)$. We separate differential structures to reconstruct
all BDC on $GL_q(N)$ possessing the usual classical limit and find the
classical counterparts for all the ingredients of BDC, {\em e.g.} we give
the"commutator" representation for the standard $\de$.
The second part is based on the paper \cite{AAM}. Here we obtain all
Poisson Lie structures on $SL(N)$ and present their quantization which
provides the quantum external algebras on $SL_q(N)$ with the classical
dimension.  We find that all these structures are nondifferential ones
that agrees with the result of \cite{FP}.

The further development of this issue for the case of other simple Lie
groups \cite{AM1,AIP} lies beyond the scope of this lecture.
\section{Graded Poisson Lie structures on the algebra of differential
forms on $GL(N)$}
\subsection{Definition}
\setcounter{equation}{0}
We start with recalling the notion of a Poisson Lie group.
The most natural way to describe it is to use the Hopf algebra terminology
\cite{Dr}.

Let $\cal{A}$ be the function algebra on a Lie group $G$
generated
by the matrix elements of a
fundamental representation $T=||t_{i}^{~j}||$ of $G$.
It is well known that $\cal{A}$ is a Hopf algebra with the coproduct
$\Delta:~\cal{A}\rightarrow\cal{A}\otimes\cal{A}$
defined on the algebra generators by:
\begin{equation}
\Delta t_{i}^{~j}=t_{i}^{~k}\otimes t_{k}^{~l}. \label{co}
\end{equation}
A group $G$ is called a Poisson Lie group if it is supplied with a
Poisson bracket compatible with $\Delta $:
\begin{equation}
\Delta \{ f,h\}_{{\cal A}}=\{\Delta f,\Delta h\}_
{{\cal A\otimes}{\cal A}}~~f,h\in {\cal A}.
\end{equation}
It is convenient to describe a Poisson structure in terms of algebra
generators $t_{i}^{~j}$. The main theorem \cite{Dr} in the theory of
Poisson Lie groups states that if $G$ is a connected semisimple
Lie group then a Poisson Lie structure is given by the
Sklyanin bracket \cite{Skl}:
\begin{equation}
\{T_1 ,T_2\}=[r,T_1 T_2 ].
\label{sk}
\end{equation}
Here we use the
standard tensor notation $T_1=T\otimes I$ and $T_2=I\otimes T$.
The matrix $r$ coming in (\ref{sk})
is a classical $r$-matrix, {\em i.e.} it
satisfies the {\em Classical Yang Baxter Equation} (CYBE) \cite{Dr}.

In this section we will treat the not semisimple General Linear Group
 ($GL(N)$), however, for the purposes of
quantization only the brackets associated with classical $r$-matrices
are relevant. Thus, following \cite{Fad} we employ (\ref{sk})
for the Poisson Lie structure on $GL(N)$
with $r$ being a trivial lifting of $r$-matrix for $SL(N)$.
It is known \cite{Ta} that the bracket (\ref{sk}) does not depend on
the $GL(N)$-invariant symmetric part of $r$-matrix.
We take $r=r_{+}$ satisfying the {\em classical Hecke condition}:
$Pr_+P+r_+=2P$,
where $P_{ik}^{~~sp}=\delta _i^{~p}\delta _k^{~s}$ is the permutation
operator.

In the following it will be important that
the bracket is degenerate and the function
$\det T$ lies in its center:
$\{ T ,\det T \}=0$.
By fixing the value of $\det T$ equal to unity we obtain the Poisson Lie
structure on $SL(N)$.

Now our goal is to extend the Poisson Lie structure
on $G=GL(N)$ to the
algebra of differential forms $\cal{M}$. (We will treat elements of
$\cal{A}$ as forms of order zero {\em i.e.} $\cal{A}\in\cal{M}$.)
The algebra $\cal{M}$ has a natural $Z_2$ grading and therefore
on homogeneous elements of $\cal M$ the corresponding
Poisson bracket should
satisfy the graded Jacobi identity,
the graded Leibniz rule,
and the graded symmetry property:
\begin{equation}
\{x, y \}=(-1)^{\deg{x}\deg{y}+1}\{y, x \} ,~~~ \deg{\{x, y
\}}=(\deg{x}+\deg{y})\bmod 2.
\label{ya2}
\end{equation}

To introduce on $\cal{M}$ the structure of a $Z_2$-graded Hopf algebra we
need to define the coproduct $\Delta$ on odd generators (one-forms) of
$\cal{M}$. The suitable set of odd generators is provided by the
matrix elements $\theta_{i}^{~j}$ of the right-invariant Maurer-Cartan form
$\Theta =\de TT^{-1}$. The reason for this choice is the well known
"gauge" transformation
law of $\Theta$ under the group translations $g\rightarrow g_1g$:
$\Theta_{g}\rightarrow \Theta_{g_1g}=g_1\Theta_{g}g_1^{-1}+dg_1g_1^{-1}$.
Hence we define the homomorphism
$\Delta :~\cal{M}\rightarrow\cal{M}\otimes\cal{M}$
on the generators $\theta_{i}^{~j}$ as:
\begin{equation}
\Delta \theta _{i}^{~j}=\theta _{i}^{~j}\otimes
I+t_{i}^{~k}S(t_{p}^{~j})\otimes \theta _{k}^{~p}.
\label{com}
\end{equation}
It is also possible to introduce
the corresponding counit and the antipod \cite{AM}. Therefore,
${\cal M}$ supplied with such a coproduct is a
genuine $Z_2$-graded Hopf algebra.
Now we are in position to give the\\[2mm]
{\bf Definition}$~~$
{\em We say that a Poisson bracket defines on ${\cal M}$ a
graded Poisson Lie structure if it is compatible with $\Delta$:
\begin{equation}
\Delta \{ x,y\}_{{\cal M}}=\{\Delta x,\Delta y\}_
{{\cal M\otimes}{\cal M}}~~x,y\in {\cal M}
\label{gpb}
\end{equation}
or, in other words, the coproduct $\Delta$ should be
a homomorphism of the Poisson algebra $\cal M$ into ${\cal M}\otimes{\cal
M}$ .} In the following we will refer to this bracket as to the
$\Delta$-covariant one.

\subsection{Explicit formulae for graded brackets}
The general expression for the bracket
involving generators $\Theta$ and $T$ ({\it bracket of the
first order}) follows from the grading requirement (\ref{ya2}):
\begin{equation}
\{\theta _{i}^{~j},t_{k}^{~l}\}=C_{ik~m}^{~~~jl~s}\theta _{s}^{~m}+
C_{ik~i_{1}i_{2}i_{3}}^{~~~jl~j_{1}j_{2}j_{3}}\theta _{j_{1}}^{~i_{1}}
\theta _{j_{2}}^{~i_{2}}\theta _{j_{3}}^{~i_{3}}+\ldots~.
\label{anz}
\end{equation}
The structure coefficients $C$ that
enter eq.(\ref{anz}) are the functions of even generators $t_{i}^{~j}$.
Applying $\Delta$ to
the both sides of (\ref{anz}) and imposing (\ref{gpb}) we obtain
the system of equations on the structure tensors $C$.
We omit the calculations of $C$ and quote only the
result. All candidates for graded $\Delta$-covariant brackets on $\cal M$
are of the form
\begin{equation}
\{\Theta _{1},T_{2}\}=r_{A}^{~12}\Theta _{1}T_{2}-\Theta
_{1}r_{B}^{~12}T_{2}+\alpha_{1} \Theta _{2}T_{2}+\alpha_{2} \tr{\Theta
}T_{2}+ \alpha_{3} \tr{\Theta }P^{12}T_{2}+\alpha_{5} \Theta _{1}T_{2}.
\label{t5}
\end{equation}
Here $r_{A}=r_{+}+\alpha _{4}P$ and
$r_{B}=r_{+}+\alpha _{6}P$.

At this stage all the numerical coefficients
$\alpha _1,\ldots, \alpha_6$ coming in (\ref{t5}) leave to be undetermined.
However, as it was mentioned above, to define the genuine Poisson
structure on $\cal M$ the bracket (\ref{t5}) should satisfy the graded
Jacobi identity.
This fixes some of the parameters
$\alpha$. It turns out that possible $\Delta$-covariant brackets are
divided into two families each of them is parametrized by two
continuous parameters $\alpha$ and $\beta$ and the discrete value of
$m=\pm 2$:
\begin{enumerate}
\item The first family
\begin{equation}
\{\Theta_1,T_2\}_{\alpha,\beta}^{\pm}=
r^{12}_{\pm}\Theta_1T_2-\Theta _1r^{12}_{\mp}T_2 +\alpha\Theta _2T_2+
\alpha_{2}\tr{\Theta}
T_2+\alpha_{3}\tr{\Theta} P^{12}T_2+\beta
\Theta_{1}T_{2}
\label {s2}
\end{equation}
where $\alpha_{2},\alpha_{3}$
are expressed via $\alpha$ as follows
\begin{equation}
\alpha_{2}=-\alpha ^2 /(m+\alpha N),~
\alpha_{3}=-\alpha m/(m+\alpha N),~~m=\pm 2, \alpha\neq -m/N.
\label{arrm}
\end{equation}
\item The second family
\begin{equation} \{\Theta_1,T_2\}_{\alpha,\beta}^{\pm}=
r^{12}_{\pm}\Theta_1T_2-\Theta _1r^{12}_{\pm}T_2 +\alpha \tr{\Theta}T_2
+\beta\Theta_{1}T_{2},
\label {s22}
\end{equation}
\end{enumerate}
where $r_{-}=r_{+}-2P$ provides another solution of the CYBE.
We clarify the geometrical meaning of $\alpha$ and $\beta$ later and
now following the same steps we define the brackets containing
two generators $\Theta$ ({\em brackets of the second order}).

We take the bracket in the general form:
\begin{equation}
\{\theta _{i}^{~j},\theta _{k}^{~l}\}=(W)_{ik}^{~jl}+W_{ik~sp}^{~jl~mn}
\theta _{m}^{~s}\theta _{n}^{~p}+\ldots ,
\label{t6}
\end{equation}
where $W$ are the unknown tensor
functions of even variables $t_i^{~j}$.
According to our general strategy we have to define all $W$-s that
guarantee the $\Delta$-covariance of a corresponding bracket.
Just at this point the crucial difference between the brackets
from the first and from the second family arises.  Namely, applying
$\Delta$ to the both sides of (\ref{t6}) and trying to obtain the system
of equation on tensors $W$ one has to use the brackets of the
first order. If one uses
the brackets  from the first family (eq.(\ref{s2})) the
corresponding system of equations for $W$ tensors can be solved and the
solution is given by
\begin{equation}
\{\Theta _1, \Theta
_2\}_{\alpha}^{\pm} =  \alpha(\Theta _1\Theta _1+ \Theta _2\Theta_2)+
r^{12}_{+}\Theta _1\Theta _2+ \Theta _1\Theta_2r^{12}_{+}- \Theta
_1r^{12}_{\mp}\Theta _2+ \Theta _2r^{12}_{\pm}\Theta _1.
\label {at13}
\end{equation}
Here the parameter $\alpha$ is the same as in (\ref{s2}) and the signs
$\pm$ are in accordance with (\ref{s2}).

As for the second family (\ref{s22}) the system of equations for $W$
tensors has no solutions and therefore there is no $\Delta$-covariant
Poisson bracket of the second order that prolongs the bracket (\ref{s22}).

One has also to check the Jacobi identity for the
system of brackets
given by (\ref{sk}), (\ref{s2}) and (\ref{at13}):
\begin{equation}
\sum (-1)^{\deg{(1)}\deg{(3)}}\{\{\Theta ,\Theta\},T\}=0~~\mbox{and}~~
\sum \{\{\Theta ,\Theta\},\Theta\}=0,
\label{j2}
\end{equation}
We omit the corresponding calculations and note only
that these equations reduce to the CYBE for $r_{\pm}$.

\subsection{Differential calculus from Poisson Lie structure}
In the previous section we have described all Poisson Lie structures
on $\cal M$ without any references to the operator of exterior derivative
$\de $. At the same time, in BDC the operator $\de  $ defined on the
function algebra on a quantum group and obeying the Leibniz rule
is one of the essential elements of the construction.
Taking the
semiclassical limit of BDC ($[a,b] \to h\{a,b\}$) we argue that it is
reasonable to introduce the following\\[2mm]
{\bf Definition}
{\em A graded Poisson structure on $\cal M$
is called a differential one if the operator
$\de $ satisfies the Leibniz-like rule:}
\begin {equation}
\de \{ f,h\}=\{{\bf  d}f,h\} +(-1)^{\deg f}\{ f, \de h\}
\label {LEB}
\vspace{2mm}
\end{equation}
In this definition the operator $\de $ is the usual differentiation
of $\cal M$ which in algebraic terms can be specified on generators
in a natural way:
\begin{equation}
\de T=\Theta T~~~\de \Theta=\Theta\Theta,
\label{ddd}
\end{equation}
and extended to the whole algebra $\cal M$ by using $\de ^2 =0$ and
the Leibniz rule.

One can see by the straightforward calculation
that the only
differential Poisson Lie structures on $\cal M$ are those for which
$\alpha=\beta$ since, {\em e.g.},
$$
\{\de T_1,T_2\}_{\alpha,\beta}^{\pm}+\{T_1,\de T_2\}_{\alpha,\beta}^{\pm}=
\de \{T_1,T_2\}+(\alpha-\beta)(\Theta_1-\Theta_2)T_1T_2.
$$

There is an alternative point of view on the operator $\de $ inspired by
the ideology of the {\em noncommutative} geometry and realized in BDC.
With Poisson Lie structures in hands we can follow this ideology
just in the classical (commutative) case. Namely,
one can define on the Poisson algebra ${\cal M}_{\alpha,\beta}^{\pm}$,
{\em i.e.} on $\cal M$ supplied with the bracket $\{,\}_{\alpha,\beta}^{\pm}$,
the operator
\begin{equation}
d_{\alpha,\beta}^{\pm} \equiv
\left\{\frac{1}{\alpha N+m}\tr{\Theta} ,~\right\}^{\pm}_{\alpha,\beta}.
\label{DT}
\end{equation}
The operator $d_{\alpha,\beta}^{\pm}$ satisfies the Leibniz rule
because the brackets
$\{,\}_{\alpha,\beta}^{\pm}$ and $\{,\}_{\alpha}^{\pm}$ do.
The property $d_{\alpha,\beta}^2=0$ is
due to the Jacobi identity and $\{ \tr{\Theta},\tr{\Theta} \}_\alpha
^\pm =0$.
Hence $d_{\alpha,\beta}^{\pm}$ is a good candidate for the
operator of exterior derivative on $\cal M$.
However, one finds:
\begin{equation}
(d_{\alpha,\beta}^{\pm}T) T^{-1}=
\Theta -\frac{\alpha-\beta}{\alpha N+m}\tr{\Theta}I,~~
d_{\alpha,\beta}^{\pm}\Theta = \Theta\Theta .
\label{DT1}
\end{equation}
Therefore, we realize again
that if and only if $\alpha=\beta$
then $d_{\alpha,\beta}^{\pm}$ can be identified with
the usual $\de $.
In other words, for graded Poisson Lie
algebras with $\alpha=\beta$ the usual operator of exterior derivative can
be expressed as the internal object of a Poisson structure.
In this case the brackets are found to be differential ones as a direct
consequence of the Jacobi identity.

It is instructive to note that in all the concrete realizations of BDC
the quantum $\de $ is defined via the graded commutator with some
tr-like differential left-right-invariant one form
\cite{J,Car,WZ}.

\subsection{Algebraic and geometric meaning of the variety of brackets}
In this section we are going to clarify the meaning of the parameters
$\alpha$ and $\beta$ that label the variety of $\Delta$-covariant
brackets on $\cal M$.
First of all let us note that there is an arbitrariness in
the choice of generators $t_i^{~j}$ and
$\theta_i^{~j}$ of $\cal M$.
There exists a nondegenerate change of variables:
\begin{equation} \label{w}
T\rightarrow \tilde{T}  =  T(\det T)^s, ~~~
\Theta \rightarrow \tilde{\Theta}  =  \Theta + k\tr{\Theta},~~
k,s\neq -1/N
\end{equation}
that does not affect the form of the coproduct:
$
\Delta \tilde{t}_i^{~j}= \tilde{t}_i^{~k}\otimes \tilde{t}_k^{~
j},~
$
$
\Delta \tilde{\theta}_{i}^{~j}=\tilde{\theta}_{i}^{~j}\otimes
I+\tilde{t}_{i}^{~k}S(\tilde{t}_{p}^{~j})\otimes \tilde{\theta}_{k}
^{~p}.
$

This covariance of the coproduct
reflects itself on the level of brackets.
Namely, one can check that the bracket (\ref{sk}) is not
affected by the transformation (\ref{w}):
\begin{equation}
\{\tilde{T}(T)_1 ,\tilde{T}(T)_2 \}=\{\tilde{T}_1 ,\tilde{T}_2 \}
|_{\tilde{T}=T\det (T)^s}.
\label{TT}
\end{equation}
This is a direct consequence of the fact that $\det T$
is the central element of the Sklyanin bracket.
However, as for the brackets of the first and the second orders
under the transformation (\ref{w}) they transform as follows
\begin{eqnarray*}
\{\tilde{\Theta}(\Theta)_1,\tilde{T}(T)_2\}_{\alpha ,\beta}^{\pm}
& = &\{\tilde{\Theta}_1,\tilde{T}_2\}_{\alpha ',\beta '}^{\pm}
|_{\tilde{\Theta}=\Theta + k\tr{\Theta},~\tilde{T}=T(\det T)^s}.
\\ \nonumber
\{\tilde{\Theta}(\Theta)_1,\tilde{\Theta}(\Theta)_2\}_{\alpha}^{\pm} &=&
\{\tilde{\Theta}_1,\tilde{\Theta}_2\}_{\alpha '}^{\pm}
|_{\tilde{\Theta}=\Theta + k\tr{\Theta}},
\label{qq}
\end{eqnarray*}
where in both cases
\begin{equation}
\alpha '=\alpha +k(\alpha N+m),~~\beta '=\beta +s(\beta N+m).
\label{AB}
\end{equation}

So, we see that under the
change of generators given by eq.(\ref{w}) the brackets $\{ ~,~\}_{\alpha
,\beta}^{\pm}$ and $\{ ~ ,~\}^\pm_\alpha$ are transformed into the brackets
$\{ ~,~\}_{\alpha ',\beta '}^{\pm}$ and $\{ ~,~\}^\pm_{\alpha '}$
with $\alpha '$ and $\beta '$ given by eq.(\ref{AB}). The inverse assertion
is also true:
for any two pairs of admissible $\alpha ,\beta$ and $\alpha ',\beta '$
($\alpha ,\beta ,\alpha ',\beta '\neq -m/N
)$ there exist such $k$ and $s$ that the transformation
(\ref{w}) converts the "$\alpha ,\beta$" brackets into the
"$\alpha ' , \beta '$" ones.

Hence, we have proved that by an appropriate change of variables any bracket
on $\cal M$ can be put into the "canonical" form with
$\alpha=0=\beta$:
\begin{eqnarray}\label{can}
\{\Theta_1,T_2\}^{\pm} & = & r^{12}_{\pm}\Theta _1T_2-\Theta
_1r^{12}_{\mp}T_2  \\ \nonumber
\{\Theta _1, \Theta _2\}^{\pm} & = & r^{12}_{+}\Theta _1\Theta _2
+ \Theta _1\Theta_2r^{12}_{+}
-\Theta _1 r^{12}_{\mp}\Theta _2+
\Theta _2r^{12}_{\pm}\Theta _1.
\end{eqnarray}

 From the geometric point of view the first equation in (\ref{w})  reflects
the existence of nonequivalent fundamental representations of $GL(N)$.
Clearly, the representation
$\hat{T} =T(\det{T})^s$
is not equivalent to $T$ for they differ by the
value of determinant:  $ \det{\hat{T}}=(\det{T})^{sN+1}.$
The point $s=-1/N$ is forbidden because the corresponding
representation becomes non-exact and can not serve as a coordinate
system on $GL(N)$.

The basis in the linear space $\Lambda^1$ of the first order differential
forms can be chosen from right(left) invariant forms in the following way.
Any representation $T$ of $G$ gives rise to the representation of the
Lie algebra $gl(N)$. For a given representation $T$ one can define the
Lie-valued right-invariant Maurer-Cartan form by:
$\Theta=\de TT^{-1}$.  The matrix elements of $\Theta$ form a basis in
$\Lambda^1$.

When one changes  a coordinate system on $G$, {\em i.e.} one
goes to some other representation $\hat{T}$, one also changes the matrix
$\Theta$. In the new basis one obtains for $\hat{\Theta}$:
\begin{eqnarray}
\hat{\Theta}=\de \hat{T}\hat{T}^{-1}=
\de \ln T+s\de \tr\ln T=\Theta+s\tr{\Theta}I.
\label{qd}
\end{eqnarray}

The expressions
for $\hat{T}$ and $\hat{\Theta}$ literally coincide with
formulae (\ref{w}) for the change of variables in $\cal M$
with the only difference: two parameters $(k,s)$ in the algebra
are replaced by the single parameter $s$ in the group.

Now we see that not every bracket with arbitrary $\alpha$ and $\beta$ can be
put in the canonical form (\ref{can}) by a change of a coordinate
system on $G$. This gives a reason to divide all Poisson Lie
structures on the external algebra into two sets. The first set
consists of the brackets with $\alpha=\beta$ and the second one with
$\alpha\neq\beta$.

Thus, we proved that there exist two different sets
of Poisson Lie structures on the external algebra of $GL(N)$.
The brackets from the first set are labeled by the signs $\pm$
and by the continuous
parameter $\alpha$, which is redundant as it can be removed
by the proper choice of a coordinate system on $G$.
These brackets are the differential ones and $\de $ operator can be
expressed as
\begin{equation}
\de ~~=\left\{\frac{1}{\alpha N \pm
2}\tr{\Theta},~~\right\}_{\alpha}^\pm.
\label{d3}
\end{equation}
The brackets from the second set are labeled by the signs $\pm$
and by a pair of parameters and
only one parameter can be set to zero by an appropriate change of
coordinates
on $G$. In contrast to the first set these brackets are not
differential ones and $\bf
d$ cannot be presented as an internal object
of this Poisson Lie structure.

\subsection{Connection with the bicovariant differential calculus on QG}
Now we can relate the graded Poisson Lie structures on $GL(N)$ with BDC
of \cite{Mu,WZ,Sudb} on  $GL_q(N)$.

BDC on $GL_q(N)$
is a free associative algebra ${\cal M}_h$
($q=e^h$) generated by the symbols
$T$ and $dT$  modulo the
quadratic relations which
in terms of quantum right-invariant forms $\Theta =dTT^{-1}$ read
\begin{eqnarray}
R_{\pm}T_{1}T_{2}&=&T_{2}T_{1}R_{\pm},\label{clr}\\
T_2 \Theta_1&=&R_\pm \Theta_1 R_\mp^{-1}T_2 \label{D1} \\
-R_\pm\Theta_1 R_\mp^{-1}\Theta_2 &=&\Theta_2 R_\pm \Theta_1 R_ \pm^{-1}.
 \label{D11}
\end{eqnarray}
where $R_+ =R$ is a quantum $R$-matrix and
$R_- =\sigma (R^{-1})$ ($\sigma$ is a permutation map) obeys the
Hecke condition:
$R_{\pm}=R_{\mp}\pm\mu P,~~\mu =q-1/q.
$

Suppose that these relations represent a quantization of a $Z_2$-graded
commutative Poisson algebra $\cal M$. Then according to the general
quantization principle one can define on $\cal M$ the Poisson bracket:
\begin{equation}
\{\Theta_1,T_2\}=-\lim_{h\rightarrow 0}\frac{1}{h}[\Theta_1,T_2]^\pm ~~
{\rm and}~~
\{\Theta_1,\Theta_2\}=-\lim_{h\rightarrow
0}\frac{1}{h}[\Theta_1,\Theta_2]^\pm
\label{bre}
\end{equation}
Here we use the square brackets $[,]$ for the graded commutator.

Since $R_\pm$ have the semiclassical form:
$ R_{\pm}=1+hr_{\pm}+o(h)~$,
the semiclassical expansion of the multiplication law in ${\cal M}_h$
reads
\begin{equation}
T_2\Theta_{1}=\Theta_{1}T_{2}+h(r_{\pm}^{12}\Theta_{1}T_2-\Theta_{1}
r_{\mp}^{12}T_2)+O(h^2),
\label{pro}
\end{equation}
\begin{equation}
\Theta_1\Theta_2=-\Theta_2\Theta_1-h(r_{\pm}^{12}\Theta_1\Theta_2+
\Theta_1\Theta_2r_{\pm}^{12}-\Theta_1r_{\mp}^{12}\Theta_2+\Theta_2
r_{\pm}^{12}\Theta_1)+O(h^2).
\label{pro1}
\end{equation}
 From the last equations we can read off at once
the expressions for the graded Poisson brackets:
\begin{equation}
\{\Theta_1,T_2\}^{\pm}=
r^{12}_{\pm}\Theta_1T_2-\Theta _1r^{12}_{\mp}T_2
\label {p2}
\end{equation}
\begin{equation}
\{\Theta _1, \Theta _2\}^{\pm} =
r^{12}_{+}\Theta _1\Theta _2+ \Theta _1\Theta_2r^{12}_{+}
- \Theta_1r^{12}_{\mp}\Theta _2+
\Theta_2r^{12}_{\pm}\Theta _1.
\end{equation}
which literally coincide with the canonical form (eq.(\ref{can})) of our
brackets.
The Diamond Condition \cite{Sudb} reduces in the semiclassical
limit to the graded Jacobi identity.

For BDC on $GL_q(N)$ the quantum operator $\de $
of exterior derivative can be
introduced in the following way \cite{Car}, \cite{WZ}. Consider
the "right-left invariant" element $\mbox{tr}_q\Theta$ called the
quantum trace of $\Theta$:
\begin{equation}\label{tr}
\mbox{tr}_q\Theta=\mbox{tr}{(D\Theta)},
\end{equation}
where $D$ is the numerical matrix  $D=diag(1,q^2,\ldots ,q^{2(N-1)})$
\cite{Fad}.

Using $\mbox{tr}_q\Theta$ let us define for the $"\pm "$ calculi the
operator
\begin{equation}
\de =\frac{1}{\gamma}\left[\mbox{tr}_q\Theta,~~ \right],
\label{tel}
\end{equation}
where $\gamma=\mu q^{2N-1}$ for the $"+"$ and  $\gamma=\mu q^{-1}$ for
the $"-"$ calculi respectively.
 From eq.(\ref{D11}) one has (see \cite{WZ} for details) the
nilpotency condition $\de ^2=0$ and the Maurer-Cartan
equation: $\de\Theta=\Theta^2$. Moreover, one also gets $\de T=dT$.
As it should be expected from the above discussion in the semiclassical
limit the quantum $\de $ reproduces eq.(\ref{d3}). In other words,
the operator $\de $ in BDC is a genuine quantization of its
classical counterpart.

To conclude this section we give a comment on quantum versions
of the noncanonical brackets (\ref{s2}) and (\ref{at13}). To obtain
these let us note that there exists the "quantum" transformation
group
\begin{equation}
T\to \tilde{T}=T(\mbox{det}_qT)^s,~~~\Theta\to \tilde{\Theta}=\Theta+
k\mbox{Tr}_q\Theta
\label{qtg}
\end{equation}
being the deformation of eq.(\ref{w}).
Since (\ref{qtg}) does not affect the coproduct the quantization of any
bracket from the family (\ref{s2}), (\ref{at13}) is obtained by
substituting $T(\tilde{T})$ and $\Theta (\tilde{\Theta })$
from eq.(\ref{qtg}) into the defining
relations (\ref{D1}), (\ref{D11}). The resulting
algebras are in agreement with the
recent classification of quantum covariant algebras on $GL_q(N)$
\cite{IS}. The transformations (\ref{qtg}) will be a useful tool in
dealing with $SL_q(N)$.

\section{Graded Poisson Lie structures on $SL(N)$ and their quantization}
\setcounter{equation}{0}
\subsection{Poisson Lie structures}
In the previous part of the lecture our aim was to establish a proper
classical structure to be the semiclassical limit of BDC on
$GL_q(N)$. To solve this task we have
introduced the notion of a Poisson Lie structure on the external algebra
on $GL(N)$. Concerning $SL_q(N)$ we have to reverse the setting of the
problem. The point is that, as we have mentioned in the Introduction, the
known algebras describing BDC on $SL_q(N)$ involve more generators as
compared to the classical case \cite{Zum,IS}. Here our goal will be
to quantize the algebra of external forms on $SL(N)$ preserving its
classical dimension. Our strategy is the
following. Among the variety of graded brackets produced in the previous
section we select the brackets which admit the $SL(N)$ reduction and present
their quantization. Carrying out this program we will be forced to modify
slightly the very definition of a graded Poisson Lie structure.

The group $SL(N)$ stands out
by the constraint $\det{T}=1$ while the external algebra on $SL(N)$ is
generated by the components $\theta_i^{~j}$
of the traceless Maurer-Cartan form.
Thus, the straightforward $SL(N)$ reduction consists
in selecting the graded Poisson Lie structures on $GL(N)$ that
are compatible with constraints
\begin{equation}
\det{T}=1~~~\mbox{and}~~~\tr{\Theta}=0.
\label{cons}
\end{equation}
However, the closer examination (see eq.(\ref{DT})) reveals that among the
graded Poisson Lie structures on $GL(N)$ given by (\ref{s2}) and
(\ref{at13}) there are no candidates to be consistent with these
constraints.

However, let us put $\tr{\Theta}=0$ in the brackets from the second family:
\begin{equation}
\{\Theta_1,T_2\}_{
\beta}^{\pm}= r^{12}_{\pm}\Theta_1T_2-\Theta _1r^{12}_{\pm}T_2
+\beta\Theta_{1}T_{2}.
\label{ll}
\end{equation}
The classical Hecke condition for $r_{+}$ guarantees that if $\beta=0$ then
$\det{T}$ and $\tr{\Theta}$ are the central elements of these brackets and,
therefore, they can be put equal to desired numerical values.  Thus, the
brackets of the first order
from the second family are the good object to admit the $SL(N)$
reduction.

Unfortunately, as we realized from the previous analysis these
brackets can not be prolonged to the $\Delta$-covariant brackets of the
second order.  It means that $\Delta$-covariance
is too strong to be satisfied for Special groups.

The hint for relaxing the constraint of $\Delta$-covariance is provided by
the Wo\-ro\-no\-wicz notion of bicovariance. Let us introduce the left and
right
coactions $\Delta_{L}$ and $\Delta_{R}$ being homomorphisms:
$\Delta_L : {\cal M}\rightarrow {\cal A}\otimes {\cal M}$,
$\Delta_R : {\cal M}\rightarrow {\cal M}\otimes {\cal A}$
defined on generators $\theta_i^{~j}$
as:
$$ \Delta_{R}\theta_{i}^{~j}=\theta_{i}^{~j}\otimes I,~~
\Delta_{L}\theta_{i}^{~j}=t_{i}^{~k}S(t_{p}^{~j})\otimes \theta _{k}^{~p}
$$
and on $t_i^{~j}$ as $\Delta_{R}=\Delta_{L}=\Delta$.
The space of
differential forms of the first order supplied with these coactions is
literally a classical counterpart of the Woronowicz bicovariant bimodule.
Hence we introduce the new\\[2mm]
{\bf Definition}$~~${\em We say that a Poisson bracket defines on ${\cal
M}$ a graded Poisson Lie structure if the following relations are
satisfied:
\begin{eqnarray}
\Delta_L \{ x,y\}_{{\cal M}} &= &\{\Delta_L
(x),\Delta_L (y)\}_ {{\cal A\otimes}{\cal M}},
\label{cc} \\
\Delta_R \{ x,y\}_{{\cal M}} &= &\{\Delta_R (x),\Delta_R (y)\}_
{{\cal M\otimes}{\cal A}}~~x,y\in {\cal M}
\label{cc1}
\end{eqnarray}
or, in other words, the coproducts $\Delta_{L,R}$ should be
homomorphisms of the Poisson algebra $\cal M$ into
${\cal A}\otimes{\cal M}$, ${\cal M}\otimes{\cal A}$ respectively.
}\\
Strictly speaking to distinguish between the requirements
(\ref{cc}), (\ref{cc1}) and
(\ref{gpb}) we have to consider
the last definition as the definition of
bicovariant Poisson Lie structure on ${\cal M}$. However for simplicity
we say Poisson Lie structure on ${\cal M}$ having in mind requirements
(\ref{cc}), (\ref{cc1}).
To avoid misunderstanding we call this bracket the {\it
bicovariant} one in contrast to the $\delta$-covariant bracket in the
definition in sect. 2.1
To justify this definition let us note that any bicovariant bracket of
the first order {\em linear} in $\theta$-s is also $\Delta$-covariant and
vice versa.

Now our goal is to define bicovariant brackets of the second order obeying
the $SL(N)$ constraints.
Solving the bicovariance conditions (\ref{cc}) and (\ref{cc1}) we get a
pair of brackets\footnote{for $N=2$ these brackets coincide owing to the
special form of the $r$-matrix}:
\begin{equation}
\{\Theta _1, \Theta
_2\}^{\pm} = \mp\frac{2}{N}(\Theta _1\Theta _1+ \Theta _2\Theta_2)+
r^{12}_{+}\Theta _1\Theta _2+ \Theta _1\Theta_2r^{12}_{+} -
\Theta_1r^{12}_{\mp}\Theta _2+ \Theta_2r^{12}_{\pm}\Theta _1.
\label{soc}
\end{equation}
One can prove that the systems of brackets (\ref{ll}) and
(\ref{soc}) satisfy the graded Jacobi identity regardless of the
sign ($\pm$) combinations, that results in four different
graded Poisson Lie structures on the external algebra
of $SL(N)$.

It is the place to state a question if these structures are the
differential ones. The direct calculation reveals that the answer is no,
{\em i.e.} the operator $\de $ does not satisfy the Leibniz-like rule.
This is a strong indication that there is no reason to expect the
quantum $\de $ to obey the Leibniz rule \cite{AAM}.
It seems quite natural because as we have seen in the case of $GL(N)$
the operator $\de $ of exterior derivative is generated by
$\tr{\Theta}$. This is a left-right-invariant form that represents the
nontrivial element from the first cohomology group $H^1(GL(N))$. However,
$H^1(SL(N))=0$ and therefore $\de $ can not be generated by an
internal element of the algebra.

\subsection{Quantization}
We start with
the quantization of the bracket of the second order eq.(\ref{soc}).
The hint is provided by the following remark.
Comparing eqs.(\ref{soc}) and (\ref{at13}) we realize that (\ref{soc})
is nothing but the bracket (\ref{at13}) with $\alpha =-\frac{m}{N}$.
But all the brackets with different
values of $\alpha$ are connected by the change of variables
eq.(\ref{AB}). The only exception is just the point $\alpha =-\frac{m}{N}$.
To reach this point from the
canonical bracket ($\alpha =0$)
we perform the transformation eq.(\ref{w}) with
$k=-1/N$ that gives the traceless $\tilde{\Theta}$.  On the other hand it
follows from eqs.(\ref{DT}) and (\ref{DT1})
that under this value of $\alpha$
$\tr{\tilde{\Theta}}$ is a central element of the bracket:
$\{\tr{\tilde{\Theta}},\tilde{\Theta}\}=0$.
Hence, the brackets of the second order can be derived from the
canonical bracket on $GL(N)$ by passing to the traceless
Maurer-Cartan form.

The necessary tools
to perform the same trick in the quantum case have been just
prepared in the previous section. These are: the defining relations
eq.(\ref{D11}) being the quantum version of the canonical bracket
on $GL(N)$ and the quantum trace eq.(\ref{tr}).

Let us decompose the form $\Theta$ in the following way
\begin{equation}
\Theta=\tilde{\Theta}+\frac{1}{\mbox{tr}D}\mbox{tr}_q \Theta.
\label{gg1}
\end{equation}
Let us stress that eq.(\ref{gg1}) should not be treated as
the usual change $\Theta \to \tilde{\Theta}$ of a basis
of external forms on $GL_q(N)$.
The constraint $\mbox{tr}_q \tilde{\Theta} =0$
makes the generators $\tilde{\theta}_i^{~j}$ linearly dependent and to
complete
a basis one has to add one more generator
$\mbox{tr}_q \Theta$. However, for $SL(N)$  $\tilde{\theta}_i^{~j}$ do
compose the basis. Moreover, substituting the decomposition (\ref{gg1})
into eq.(\ref{D11}) we see that generators $\tilde{\Theta}$ form the
closed subalgebra. Namely,
\begin{equation}
R_{\pm}\tilde{\Theta}_1R_{\mp}^{-1}\tilde{\Theta}_2+\tilde{\Theta}_2R_{\pm}
\tilde{\Theta}_1 R_{\pm}^{-1}=
-\frac{1}{\mbox{tr}D}
\left(R_{\pm}[\mbox{tr}_q\Theta,\tilde{\Theta}_1]R_{\mp}^{-1}+
[\mbox{tr}_q\Theta,\tilde{\Theta}_2]\right).
\label{gg2}
\end{equation}
where we have used $(\mbox{tr}_q\Theta)^2=0$  and
the Hecke condition for $R_{\pm}$.
Using the Maurer-Cartan equation  for $\Theta$ one can easily find
the commutator for $\mbox{tr}_q\Theta$ and $\tilde{\Theta}$:
\begin{equation}
\left[\mbox{tr}_q\Theta, \tilde{\Theta}\right]=
\frac{\gamma \mbox{tr}D}{\mbox{tr}D-\gamma}\tilde{\Theta}^2.
\label{gg3}
\end{equation}
Taking into account the last equation
we can write down eq.(\ref{gg2}) as
\begin{equation}
R_{\pm}\tilde{\Theta}_1R_{\mp}^{-1}\tilde{\Theta}_2+\tilde{\Theta}_2R_{\pm}
\tilde{\Theta}_1 R_{\pm}^{-1}=
k_{q^{\pm 1}}
\left(R_{\pm}\tilde{\Theta}_1^2R_{\mp}^{-1}+
\tilde{\Theta}_2^2\right).
\label{gg4}
\end{equation}
where $k_q=\frac{q^N\mu}{q^N\mu+[N]_q}$ and $[N]_q$ is the q-number:
$[N]_q=(q^N-q^{-N})/\mu$.
Now one can easily verify that the formulae (\ref{gg4}) provide the
quantization of the bracket (\ref{soc}). This formula was proposed in
\cite{IS} as commutation relations for Cartan 1-forms on $SL_q(N)$.

The quantization of the brackets of the first order is
obviously given by
\begin{equation}
T_2\tilde{\Theta}_1=R_{\pm}\tilde{\Theta}_1R_{\pm}^{-1}T_2.
\label{gg5}
\end{equation}
The algebras with these relations have two central elements:
$\mbox{tr}_q\tilde{\Theta}$ and $\mbox{det}_q T$.
Combining eqs.(\ref{gg4}) and (\ref{gg5}) we get the
quantization of the graded Poisson Lie structure on $SL(N)$.
In the recent paper \cite{FP} the algebras (\ref{gg4}) and (\ref{gg5})
appeared in a pure quantum treatment.

$$~$$
{\bf ACKNOWLEDGMENT}
One of
the authors (G.A.) is grateful to the organizers of the XXX Karpacz Winter
School of Theoretical Physics for their hospitality and to
for the partial financial support of the participation in the XXX Karpacz
Winter School.
The authors are also grateful to
J.Lukierski, P.P.Kulish, M.L.Ge, A.P.Isaev and P.N.Pyatov for interesting
discussions.
This work is supported in part by RFFR under grant N93-011-147 and ISF
under grant M1L-000.

\end{document}